\documentclass[twocolumn,showpacs,preprintnumbers,amsmath,amssymb]{revtex4}

\usepackage{graphicx}
\usepackage{dcolumn}
\usepackage{bm}

\begin{document}

\preprint{APS/123-QED}

\title{Current-induced break of inversion symmetry  in Si: optical second-harmonic generation induced by a dc current}

\author{O.A. Aktsipetrov}%
\email{aktsip@shg.ru;   http://www.shg.ru}
\author{V.O. Bessonov}%
\author{A.A. Fedyanin}%
\author{A.I. Maidykovsky}%
\affiliation{%
Department of Physics, Moscow State University, 119992 Moscow,
Russia}%

\date{\today}

\begin{abstract}
{The dc-current-induced optical second-harmonic generation is
observed at the surface of centrosymmetric single crystal of Si.
Surface dc-current with density up to $\textit{j}_{max} \sim 10^{3}$
A/cm$^{2}$ results in break of inversion symmetry of Si and induces
optical second-harmonic generation with intensity that corresponds
to the appearance of dipole second-order susceptibility
$\chi^{(2)d}(j_{max})\sim  3 \cdot 10^{-15}$m/V.}
\end{abstract}

\maketitle

Optical second-harmonic generation (SHG) in reflection from the
surface of centrosymmetric materials is attractive topic of studies
from the very beginning of nonlinear optics and dates back to the
1968's experimental paper by N. Bloembergen and coworkers [1]. For
this long-term period of studies and applications SHG was recognized
as versatile powerful probe of surfaces and nanostructures [2]. High
surface and interface sensitivity of SHG was a base of the broad
family of nonlinear optical characterizations. Important role in
these probes belongs to the studies of SHG under external impacts.
The dc-electric-field-induced SHG (EFISH) [3] and
dc-magnetic-field-induced SHG (MSHG) [4] are broadly used nowadays
for characterization of electron and magnetic properties of surfaces
and nanostructures.

Silicon is one of the materials thoroughly studied be means of the
surface SHG. Apart from the general interest to this semiconductor,
Si serves as basic model of the centrosymmetric material for
nonlinear optical studies. Centrosymmetric materials are very
specific in nonlinear optics as all even-order nonlinear
susceptibilities vanish away in this class of materials [5]. As a
consequence, SHG is strongly forbidden in the balk of
centrosymmetric materials in dipole approximation due to symmetry
selection rules. Anyway, the weak second-order susceptibility of the
bulk still exists and is attributed to the quadrupole contribution.
This quadrupole susceptibility is responsible for a weak background
SHG contribution originating from the bulk. Basically this
quadrupole SHG is the sole second-order nonlinear response in
infinite samples of centrosymmetric materials.

Situation is significantly changed in semi-infinite samples: quite
strong dipole surface SHG appears. There are three major mechanisms
of the break of inversion symmetry at interfaces of centrosymmetric
materials:

1. Inversion symmetry is broken at the boundary of semi-infinite
sample due to the discontinuity of the crystallographic structure at
the interface [6]. The thickness of surface layer with broken
symmetry can be estimated as several periods of crystal lattice. The
interfacial break of inversion symmetry results in appearance of
dipole surface contribution that is dominant in reflected SHG.

2. Another mechanism of dipole reflected SHG is related to
subsurface dc-electric field in space charge region (SCR) of
semiconductors that appears in the vicinity of interfaces due to the
band banding effect. A dc electric field in SCR, that is normal to
the surface, breaks inversion symmetry in SCR and provides the
strong EFISH contribution [7]. The thickness of SCR varies from
several nanometers to a few hundreds of nanometers that depends on
band banding potential and doping of semiconductor.

3. Third mechanism is related to the nonhomogeneous surface stress
that also breaks inversion symmetry at interface and provides dipole
susceptibility of surface layer and dipole reflected SHG [8]. The
thickness of disturbed layer depends on stress relaxation and is in
the range of several nanometers.

The corresponding nonlinear polarization of semi-infinite
centrosymmetric crystal is given by the following formal equation:
\begin{equation}
\begin{array}{l}
\textbf{P}_{2\omega}\propto[\widehat{\chi}^{(2)q,b}(\textbf{k}) +
\widehat{\chi}^{(2)d,s} + \widehat{\chi}^{(2)d,b}(E_{0}) + \\
\widehat{\chi}^{(2)d,s}(\sigma)]\textbf{E}_{\omega}\textbf{E}_{\omega},
\end{array}
\end{equation}
where $\widehat{\chi}^{(2)q,b}(\textbf{k})$ is quadrupole bulk
susceptibility that originates from spatial dispersion and depends
on wave vector $\textbf{k}$ of fundamental field
$\textbf{E}_{\omega}$, $\widehat{\chi}^{(2)d,s}$ is surface dipole
susceptibility that originates from the break of the crystalline
structure at the boundary of semi-infinite sample,
$\widehat{\chi}^{(2)d,b}(E_{0})$ is dipole susceptibility of SCR
that appears from the break of inversion symmetry by dc electric
field $E_{0}$ and $\widehat{\chi}^{(2)d,s}(\sigma)$ is dipole
surface susceptibility induced by strain which is described by
strain tensor $\widehat{\sigma}$. For the last decades a number of
experimental and theoretical works were devoted to the studies of
these mechanisms of surface SHG in metals and semiconductors [9-11].

Generally, all these contributions to the SHG are attributed to the
noncentrosymmetric distortion of the structure of the surface
crystallographic unit cell. Initially centrosymmetric unit cell
becomes noncentrosymmetric either due to existence of surface
boundary or appearance of strong dc electric field due to band
bending and inhomogeneous surface stress in surface layer, e.g. due
to lattice missmatch between Si and SiO$_{2}$.
\begin{figure}[!h] \vspace{0.1 cm}
\begin{centering}
\includegraphics
[width=5.5cm]{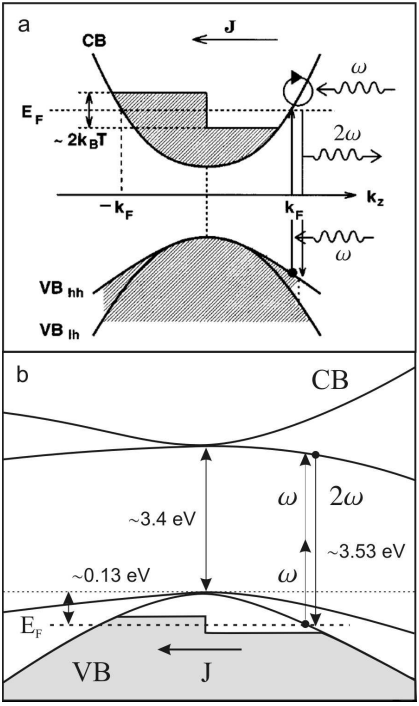} \caption{Panel a: schematic view of band
structure of model direct gap n-doped semiconductor (from Ref.
[12]). Current $J$ disturbs electron distribution function in the
vicinity of local Fermi level $E_{F}$ that is schematically shown
by step-like feature in conduction band.  Arrows show electron
transitions (real or virtual) assisted with fundamental and SHG
photons. Panel b: the same scheme for $p$-doped Si.}\label{1}
\end{centering}
\vspace{0.1 cm}
\end{figure}
Meanwhile, apart from these contributions conditioned by
noncentrosymmetric distortions of crystalline lattice one more
mechanism of the break of inversion symmetry exists that is almost
missed so far. The dc current flowing through semiconductor distorts
equilibrium distribution function of electrons in quasi-impulse
domain. This breaks inversion symmetry of electron subsystem of
centrosymmetric semiconductor and results in appearance of the new
term in Eq. 1:
$\textbf{P}_{2\omega}(\textbf{j})=\widehat{\chi}^{(2)d}(\textbf{j})\textbf{E}_{\omega}\textbf{E}_{\omega}$,
where $\textbf{j}$ is current density and
$\widehat{\chi}^{(2)d}(\textbf{j})$ is corresponding current-induced
susceptibility. Theory of dc-current-induced SHG (CISH) in model
direct band \textit{n}-doped semiconductor is developed by Khurgin
in Ref. [12]. Figure 1a shows schematic view of band structure of
direct band semiconductor with electron distribution function in
conduction band disturbed by a dc current (from Ref. [12]).
Distribution of electrons is asymmetric for $\textbf{k}_{el}$ and
$-\textbf{k}_{el}$ , where $\textbf{k}_{el}$ is electron
quasi-impulse. Perturbation theory approach results in appearance of
current-induced term in second-order susceptibility with sharp
resonance in the vicinity of local Fermi level for majority carriers
in conduction band, $E_{F}$. From theoretical consideration it
follows that $\widehat{\chi}^{(2)d}(\textbf{j})\propto j$, where
$j=\mid \textbf{j}\mid$ and
$\widehat{\chi}^{(2)d}(\textbf{j})=-\widehat{\chi}^{(2)d}(-\textbf{j})$
that implies that current should break inversion symmetry of
centrosymmetric materials. Symmetry analysis shows that for s-in,
s-out combination of polarization of the fundamental and SH waves,
respectively, the CISH signal appears as $\textbf{j}$ is normal to
the plane of incidence: for transversal geometry, and vanishes as
$\textbf{j}$ is in plane of incidence: for longitudinal geometry.

In this Letter current-induced break of the inversion symmetry and
dc-current-induced SHG are observed  in centrosymmetric Si.
Three-fouled objective is targeted in these studies: relation
between symmetry properties of electron subsystem and transport
phenomenon in semiconductor, new nonlinear optical effect in
nonequilibrium semiconductor, new probe for characterization and
imaging of surface current distribution at semiconductor surfaces
and interfaces.

There are the following points in realization of this experiment:

1. Fabrication of silicon structure that allows to pass dc current
with density high enough to induce noncentrosymmetric perturbation
of electron distribution function.

2. Selection of experimental conditions to avoid the influence of
all other contributions except $\widehat{\chi}^{(2)d}(\textbf{j})$.

3. Selection of experimental conditions to avoid artefacts related
to heating the samples by current.

4. The direct CISH effect should be distinguished from the in-plane
EFISH effect, i.e. SHG induced by the dc electric field $E_{driv}$,
that drives in-plane dc current.

Figure 2a shows schematic view of Si structure for observation of
the CISH effect. Highly \textit{p}-doped (B-doped, $\rho\sim
10^{-3}\Omega $cm) Si(001) wafer is used as substrate for
experimental structure. Square-shaped (approximately of 0.5
cm$\times$0.5 cm area) Ni-electrodes are deposited on top of Si(001)
wafer with native oxide by thermal evaporation of Ni in residual
vacuum $\sim$ $10^{-6}$ torr. The gap between Ni-electrodes is
oriented along Y crystallographic axis. The width of the gap is
200$\pm$20 microns. The thickness of Ni film is 300$\pm$20 nm. After
deposition of Ni stripes the samples were annealed to form Ohmic
contacts between Si wafer and Ni electrode in accordance with
procedure described elsewhere [13]. Contacts exhibit linear $I-V$
dependence and Ohmic resistance of approximately 0.02 $\Omega$.
Maximum voltage applied to electrodes without noticeable heating is
up to 0.5 V. The samples are mounted on special cooler stage to
reduce the heating. Directly measured temperature of the sample
during nonlinear optical experiments is less than 40$^{\circ}$C.
Estimated current density in subsurface layer with the thickness of
$\sim$ 50 nm, that corresponds to the escape depth of the SHG
radiation at wavelength $\lambda_{2\omega}$= 390 nm, is
approximately $\textit{j}_{max} \simeq 10^{3} A/cm^{2}$.

The output of an unamplified Ti:sapphire laser with a tuning range
of wavelength from 710 nm to 850 nm, a pulse duration of 80 fs,
spectral bandwidth of 10 nm, an average power of 130 mW and
repetition rate of 86 MHz is used as the fundamental radiation in
nonlinear optical studies. The train of femtosecond pulses at a 45°
angle of incidence is focused onto the Si(001) surface in the 200
micron gap between Ni-electrodes in sport with diameter of 40
microns.The SHG radiation is filtered out by appropriate glass
bandpass and interference filters and is detected by a
photo-multiplier tube and a gated photon-counting electronics. A
fraction of the fundamental laser beam was split off to generate a
SH reference signal from a crystalline Z-cut quartz plate, which
normalized against drifts in average laser power and pulse duration
during data acquisition. A detection system in "reference" arm is
identical to that in the "sample" arm.

To distinguish between current-induced and heating-induced effects
this is worth noting that current-induced susceptibility changes its
sign under reversal of current direction. This implies that the
nonlinear optical probe of the CISH effect should be sensitive to
the current direction. The dc-current-induced SHG is studied by
means of the SHG interferometry [14] with external homodyne
reference [15]. Figure 2b shows the experimental scheme of the
nonlinear optical interferometry. A 30 nm thick indium-tin-oxide
(ITO) film on a glass substrate is used as homodyne reference.

The total SHG intensity from the sample and the reference depends on
current density, current direction, the reference displacement, $r$,
and is given by:
\begin{equation}
\begin{array}{l}
I_{2\omega}^{\pm}(j,r)\propto\mid\textbf{E}_{2\omega}^{ref}+
\textbf{E}_{2\omega}^{samp}(\textbf{j})\mid^{2}=
(E_{2\omega}^{ref})^{2}+(E_{2\omega}^{samp}(j))^{2}\pm \\
2\alpha (E_{2\omega}^{ref}E_{2\omega}^{samp}(j))\cos [\displaystyle
\frac{2\pi r}{L}+\Phi^{ref}+\Phi^{samp}],\label{1}
\end{array}
\end{equation}
where
$\textbf{E}_{2\omega}^{samp}(\textbf{j})=E_{2\omega}^{samp}(j)\exp(i\Phi^{samp})$
and $\textbf{E}_{2\omega}^{ref}=E_{2\omega}^{ref}\exp(i\Phi^{ref})$
are complex amplitudes of the current-reversal and
current-independent SH fields from sample and reference,
respectively; $E_{2\omega}^{samp}(j)$, $E_{2\omega}^{ref}$,
$\Phi^{samp}$ and $\Phi^{ref}$ are real amplitudes and phases of SH
fields, respectively;  $\alpha$ is the coherence coefficient of the
fundamental beam, $L=\lambda_{\omega}(2\Delta n)^{-1}$ is period of
interference pattern and $\Delta n=n(2\omega)-n(\omega)$ describes
the dispersion of the refractive index  of air, $n$, at the SH and
fundamental wavelengths, respectively. Interference of two
components of the SH field results in appearance of homodyne
cross-term that changes its sign under current reversal and
oscillates as a harmonic function of the reference displacement. The
dc-current-induced SHG is characterized by the CISH contrast, which
is given by:
\begin{equation}
\begin{array}{l}
\rho_{j}=\frac{I_{2\omega}^{+}(j,r)-I_{2\omega}^{-}(j,r)}{I_{2\omega}^{ref}}
\propto
\\4E_{2\omega}^{ref}E_{2\omega}^{samp}(j)\cos [\displaystyle
\frac{2\pi r}{L}+\Phi^{ref}+\Phi^{samp}],\label{1}
\end{array}
\end{equation}
To avoid the influence of current sensitive variations of the SHG
intensity related to EFISH and the stress-induced SHG the azimuthal
anisotropic properties of SHG from Si(001) are used. From Refs. [6]
it follows that for s-in,s-out combination of polarizations of the
SH and fundamental waves the anisotropic SH response from Si(001)
originates only from quadrupole bulk term
$\widehat{\chi}^{(2)q,b}(\textbf{k})$ whereas other terms from Eq. 1
vanish away: these terms contribute to SHG as p-polarized component
exists either for SH or fundamental fields. Moreover, one can get
rid of quadrupole bulk contribution for proper mutual azimuthal
orientation of Si(001) wafer and incident plane of fundamental
radiation. Figure 1c shows dependence of intensity of the bulk
quadrupole SHG on the azimuthal angle for the s-in,s-out
combination, $I_{2\omega}^{s,s}(\theta)$, that demonstrates
eight-fold symmetry. If plane of incidence is set at the angle
$\theta_{zero}$ that corresponds to zero of $I_{2\omega}^{s,s}$,
this gets rid of the last contribution allowed for s-in, s-out
combination. It means that in these experimental conditions the SHG
intensity gets its zero value within the experimental error bar and
one can expect observation of the CISH signal on this "zero
background".

Thus, at "zero background" conditions of interferometric experiment
in transversal geometry of electrical bias we expect appearance of
the CISH signal that should be sensitive to reversal of $\textbf{j}$
and linear with respect to $j$ as this is proportional to
interferometric cross-term in Eq. 2, and vanishes under transition
to longitudinal geometry.
\begin{figure}[!h] \vspace{0.1 cm}
\begin{centering}
\includegraphics
[width=5.5cm]{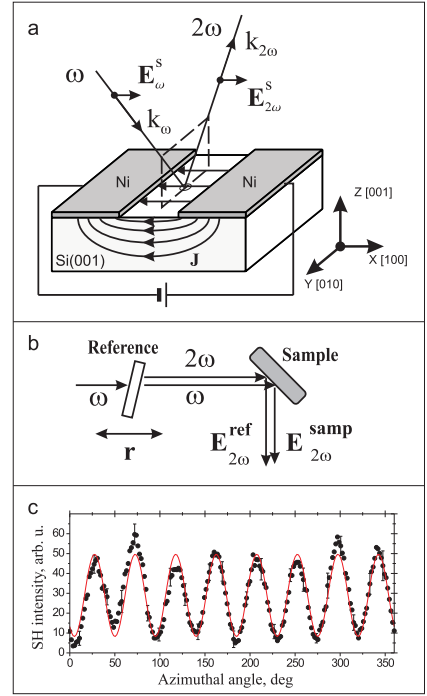} \caption{Panel a: the schematic view of Si
surface structure with Ni electrodes and details of nonlinear
optical experiment; $k_{\omega}$, $k_{2\omega}$, $E_{\omega}^{s}$
and $E_{2\omega}^{s}$ are wave vectors and $s$-polarized fields of
fundamental and SH waves, respectively. Coordinate frame
corresponds to crystallographic orientation of Si wafer. Panel b:
the scheme of the SHG interferometry. Panel c: azimuthal
anisotropic dependence of the SHG intensity for $s$-in, $s$-out
combination of polarizations.}\label{1}
\end{centering}
\vspace{0.1 cm}
\end{figure}
It turns out in experiment, the current reversible CISH signal
appears in  the external homodyne SHG interferometry for s-in,s-out
combination and transversal geometry: Figure 3a shows dependence of
$\rho_{j}$ on reference displacement. The solid line is result of
approximation by oscillatory part of Eq. 3 with $L$= 4.8 cm that is
in good agreement with $\Delta n$ at wavelength of 780 nm [16]. The
interferometric scheme allows to maximize $\rho_{j}$ value: further
measurement are performed at $r_{max}$ corresponding to the maximum
on dependence of $\rho_{j}$ on $r$. The analogous measurements of
$\rho_{j}$ are performed for s-in,s-out combination and longitudinal
geometry. Figure 3a demonstrates the lack of the CISH effect for the
longitudinal geometry: the detected SHG signal remains be equal to
"zero background" within the experimental error bar.
\begin{figure}[!h] \vspace{0.1 cm}
\begin{centering}
\includegraphics
[width=6.0cm]{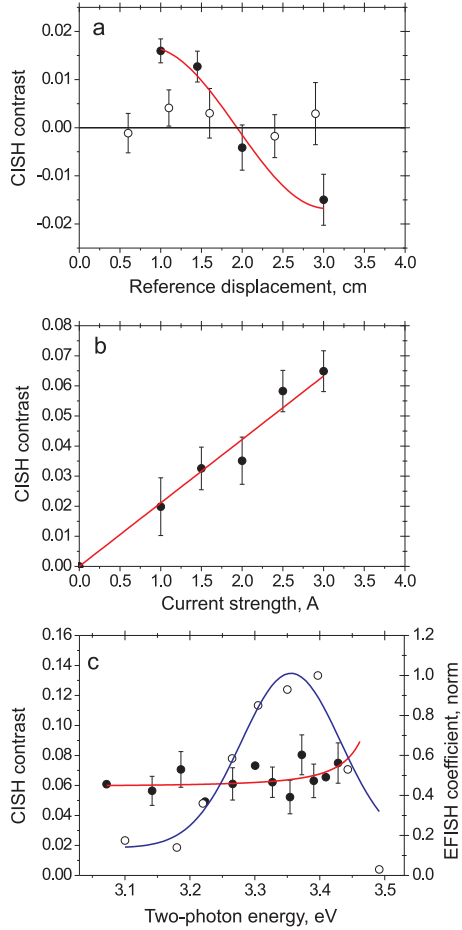} \caption{Panel a: interferometric
dependence of CISH contrast at fundamental wavelength of 780 nm
for transversal geometry, filled symbols (measured at $J$ =1 $A$),
and for longitudinal geometry, open symbols (measured at $J$ =4
$A$). The latter show the lack of CISH effect. Panel b: current
dependence of CISH signal. Panel c: spectral dependence of CISH
effect, filled symbols (measured at $J$ =4 $A$), and spectral
dependence of EFISH effect, open symbols (from Ref. [18]); solid
lines are guide for eye. }\label{1}
\end{centering}
\vspace{0.1 cm}
\end{figure}
Figure 3b shows linear dependence of $\rho_{j}$ on $j$ at fixed
position of reference with respect to the sample, which corresponds
to the measurements of current dependence of
$\widehat{\chi}^{(2)d}(\textbf{j})$ in accordance to Eq. 3. The
linear current dependence of $\rho_{j}$ reveals analogous dependence
of interference cross-term in Eq. 3 and as a consequence linear
dependence of $\widehat{\chi}^{(2)d}(\textbf{j})$ on $j$.

There is the only one source of the SHG signal, which can disguise
the CISH effect in our experimental conditions. This is in-plane
EFISH. Two experimental arguments prove that the SHG signal that
accompanies a dc current corresponds to the direct CISH effect
instead of the in-plane EFISH.

The first argument is comparison of the expected value of the
in-plane EFISH signal $I_{2\omega}(E_{driv})$ from our in-plane
biased Si and EFISH signal $I_{2\omega}(E_{SCR})$ from space charge
region of Si in biased metal-oxide-semiconductor (MOS) capacitors.
The ratio of these EFISH signals is proportional to the squared
ratio of the corresponding dc electric fields:
$I_{2\omega}(E_{driv})/I_{2\omega}(E_{SCR})\propto[E_{driv}/E_{SCR}]^{2}$
[17]. The value of the in-plane dc electric field $E_{driv}= 10^{3}$
V/m is typical for our Si structure and experimental conditions.
Whereas typical value of the dc electric field normal to the surface
is $E_{SCR}= 10^{7}$ V/m [18]. Expected ratio of EFISH signals is
$I_{2\omega}(E_{driv})/I_{2\omega}(E_{SCR})\sim10^{-8}\div10^{-9}$.
Meanwhile, the absolute value of the CISH contribution to the SHG
intensity $I_{2\omega}(j)$ can be recalculated from $\rho_{j}$. The
corresponding ratio $I_{2\omega}(j)/I_{2\omega}(E_{SCR})\sim 5
\cdot10^{-6}$ that is by two orders of magnitude large than for the
estimated in-plane EFISH contribution.

Another argument is comparison of spectral dependence of the CISH
and EFISH effects in p-Si(001). Figure 3c shows spectral
dependencies of the CISH contrast measured in our experiments and
the EFISH coefficient from Ref. [18]. Comparison of these spectra
shows the difference that prove our conclusion that observed
current-induced variations of the SHG intensity are not related to
the EFISH effect. Moreover, spectral dependence of the CISH signal
is in qualitative agreement with results of model consideration in
Ref. [12]. Figure 1b shows schematically the band structure of Si.
In the case of $p$-doped Si distribution function for holes is
similar to that for electrons considered in Ref. [12]. Local Fermy
level for holes at room temperature for our highly $p$-doped Si
wafers is 0.13$\pm$0.01 eV below the bottom of valance band at
$k$=0. This implies that sharp CISH resonance (Khurgin resonance) is
expected approximately at 3.53 eV whereas the SHG spectroscopy in
Figure 3c cover the spectral range below 3.4 eV that is restricted
by the tuning range of Ti:sapphire laser. Slightly arising spectral
dependence of the CISH contrast in Figure 3c can be qualitatively
associated with the low-energy wing of the temperature broadened
Khurgin resonance.

Comparison of the CISH signal with reflected SHG from crystalline
quartz which dipole second-order susceptibility is known from
handbooks [19] allows to estimate maximum magnitude of the CISH
susceptibility $\chi^{(2)d}(j_{max})\sim  3 \cdot 10^{-15}$m/V and
the CISH coefficient $\beta\sim 2 \cdot 10^{-8}$ m$^{2}$/A, which is
defined by: $\widehat{\chi}^{(2)d}(j)=\beta \cdot j$.

In conclusion, the dc-current-induced optical SHG is observed in
centrosymmetric single crystal of Si.  dc current with surface
density up to $\textit{j}_{max} \sim  10^{3}$ A/cm$^{2}$ induces
optical SHG with intensity that corresponds to the second-order
susceptibility $\chi^{(2)d}(j_{max})\sim  3 \cdot 10^{-15}$m/V.
Details of the CISH experiment: azimuthal orientation of Si(001)
surface, polarization combination, the SHG interferometry and
spectroscopy techniques, prove the mechanism of the SHG current
dependence that is current-induced break of inversion symmetry of
Si. Observation of the CISH effect opens perspectives of this novel
surface probe in characterization of semiconductor devices: surface
current imaging and mapping.


\begin{thebibliography}{99}
\bibitem {c1} N. Bloembergen, R.K. Chang, S.S. Jha, Phys. Rev.,
\textbf{174}, 813 (1968).

\bibitem {c2} Y.R. Shen, Nature (London), \textbf{337}, 519 (1989).

\bibitem {c3} S.H. Lee, R.K. Chang, N. Bloembergen, Phys. Rev. Lett.
\textbf{18}, 167 (1967).

\bibitem {c4} Ru-Pin Pan, H.D. Wei, Y.R. Shen, Phys. Rev. B, \textbf{39},
1229(1989).

\bibitem {c5} Y.R. Shen, \textit{The Principles of Nonlinear Optics}.
(Wiley, New York, 1984).

\bibitem {c6} H.W.K. Tom, T.F. Heinz, Y.R. Shen, Phys. Rev. Lett.
\textbf{51}, 1983 (1983).

\bibitem {c7} J.I. Dadap, X.F. Hu, M.H. Anderson, M.C. Downer, J.K. Lowell, O.A.
Aktsipetrov, Phys. Rev. B \textbf{53}, 7607R (1996).

\bibitem {c8} W. Daum, H.-J. Krause, U. Reichel, H. Ibach, Phys. Rev. Lett. \textbf{71}, 1234
(1993).

\bibitem {c9} T.F. Heinz, in  \textit{Nonlinear Surface Electromagnetic Phenomena},
H.-E. Ponath and G.I. Stegeman, eds. (North Holland, Amsterdam,
1991).

\bibitem {c10} J.F. McGilp, Phys. Stat. Sol. A \textbf{175}, 153(1999) .

\bibitem {c11} G. L\"{u}pke, Surf. Sci. Rep. \textbf{35}, 75 (1999).

\bibitem {c12} J.B. Khurgin, Appl. Phys. Lett. \textbf{67}, 1113
(1995).

\bibitem{c13} A. Singh, W.S. Khokle,   Proceedings of the
IEEE \textbf{75},  852 (1987).


\bibitem {c14} G. Berkovic, Y.R. Shen, G. Marowsky, and R. Steinhoff, J. Opt. Soc.
Am. B \textbf{6}, 205 (1989).


\bibitem{c15} J.I. Dadap, J. Shan, A.S. Weling, J.A. Misewich, A. Nahata,
T.F. Heinz, Opt. Lett. \textbf{24}, 1059 (1999).

\bibitem{c16} O.A. Aktsipetrov, E.D. Mishina, T.V. Misuryaev, A.A. Nikulin,
V.R. Novak, R. Stolle, and Th. Rasing, Surf. Sci. \textbf{402-404},
576 (1998).

\bibitem{c17} O.A. Aktsipetrov, A.A. Fedyanin, V.N. Golovkina, T.V.
Murzina,
Opt. Lett. \textbf{19}, 1450 (1994).

\bibitem{c18} O.A. Aktsipetrov,  et al. Phys. Rev. B, \textbf{60}, 8924 (1999).

\bibitem{c19} R.J. Pressley, edit. \textit{Handbook of lasers with selected
data on optical technology}. (Chemical Rubber Co. Cleveland, 1971).









\end{thebibliography}
\end{document}